\documentclass[a4paper,12pt]{scrartcl}
\usepackage{enumerate}
\usepackage[english]{babel}
\usepackage[disable]{todonotes}
\newenvironment{itquote}
{\begin{quote}\itshape}
{\end{quote}}

\usepackage[a4paper,top=2cm,bottom=2cm,left=3cm,right=3cm,marginparwidth=1.75cm]{geometry}

\newcommand{\pjs}[1]{}
\newcommand{\mb}[1]{}
\newcommand{\vt}[1]{}
\newcommand{\dv}[1]{}
\newcommand{\pbs}[1]{}
\newcommand{\commentout}[1]{}

\usepackage{amsmath}
\usepackage{amsfonts}
\usepackage{amssymb}
\usepackage{booktabs}
\usepackage{graphicx}

\usepackage[colorlinks=true, allcolors=blue]{hyperref}

\addto\extrasenglish{

}

\DeclareMathOperator{\Bi}{Bi}

\title{Assessing the accuracy of the Australian Senate count}
\subtitle{Key steps for a rigorous and transparent audit}

\author{ \normalsize
Michelle Blom \\
\normalsize \href{mailto:michelle.blom@unimelb.edu.au}{\nolinkurl{michelle.blom@unimelb.edu.au}}
\and 
\normalsize Philip B.\ Stark \\
\normalsize \href{mailto:michelle.blom@unimelb.edu.au}{\nolinkurl{stark@stat.berkeley.edu}}
\and 
\normalsize Peter J.\ Stuckey \\
\normalsize \href{mailto:michelle.blom@unimelb.edu.au}{\nolinkurl{peter.stuckey@monash.edu}}
\and 
\normalsize Vanessa Teague \\
\normalsize \href{mailto:michelle.blom@unimelb.edu.au}{\nolinkurl{vanessa.teague@anu.edu.au}}
\and 
\normalsize Damjan Vukcevic \\
\normalsize \href{mailto:damjan.vukcevic@unimelb.edu.au}{\nolinkurl{damjan.vukcevic@unimelb.edu.au}} 
}

\begin{document}
\maketitle

\section{Introduction}
Australian paper-based elections counted in a polling place are carefully designed to be both privacy-preserving and verifiable. Scrutineers can watch each voter put a ballot paper in each box (one for the Senate and one for the House of Representatives), then watch the manual count of first preferences, or the two-candidate-preferred count, after the polls close. The counting of postal votes can also be scrutinised, although postal voting entails considerable trust in the postal service.

Senate votes are electronically counted after the ballot papers are scanned and the scans transformed into digital preferences in a hybrid human-automated process.
The  count itself is easy to check, because the Australian Electoral Commission (AEC) publishes detailed digitised preferences, and there are several
open-source Senate counting implementations that allow anyone to redo the count.\footnote{\url{https://github.com/AndrewConway/ConcreteSTV}} \footnote{\url{https://github.com/grahame/dividebatur2}} \footnote{\url{https://yingtongli.me/git/OpenTally/}} \footnote{\url{https://github.com/MiladKetabGhale/STV-Counting-ProtocolVerification}}

This paper is therefore not about verifying the count, but about auditing  the scanning and digitisation of the Senate ballot papers. 
Although scrutineers are allowed to observe the automated scanning and partially automated digitisation process, there remains an important evidence gap:  scanners are trusted to produce an accurate image of each ballot, software is trusted to retain the images correctly, the digitisation process is trusted to properly digitise, and the digital preferences are assumed to be accurately output, but scrutineers cannot meaningfully observe whether any of this is done correctly. 
The process can fail in many ways.
Images could be inaccurate.
Characters could be misinterpreted (by humans or computers).
And incorrect or insecure recording and storage of preference data could alter, delete, or add preferences.
Thus, undetected software errors or security problems may cause the digitised preferences to differ from the paper ballots.
For example, bugs have been discovered in scanners that alter numbers in documents.\footnote{\url{https://www.dkriesel.com/en/blog/2013/0802_xerox-workcentres_are_switching_written_numbers_when_scanning}}  
Even random errors may not be politically neutral~\cite{evote2020errors}.

This risk will be significantly reduced by the recently passed 
\textit{Electoral Legislation Amendment (Assurance of Senate Counting) Bill 2021}.
The bill mandates an audit of the ballot papers, to compare them with the corresponding digitised preferences and establish an estimate of the overall error rate. It is significant that the law specifies an audit of ballot \emph{papers}, so that even problems associated with imperfect or insecure scanning can be detected. This is a tremendous opportunity to earn public trust by providing evidence of an accurate process.

Of course, the 
audit has to be carefully designed to allow for genuine scrutiny. 
In this guide, we give a short summary of the motivation for auditing paper ballots, explain the necessary supporting steps for a rigorous and transparent audit, and suggest some statistical methods that would be appropriate for the Australian Senate. Our aim is to fill in the AEC's published methodology sketch\footnote{\url{https://www.aec.gov.au/About_AEC/cea-notices/files/2022/s273AC-senate-assurance-methodology-fe2022.pdf}} with some more of the relevant statistical and computational details.

\textbf{Update:} This version (22 June 2022) now includes analysis of the Senate preference data for the 2022 Australian election.
See Section~\ref{subsubsec:2022Data}.

\subsection{What is auditing?}
The aim of an election audit is to ascertain whether the election result was altered by errors, regardless of their cause.  
The audit should conclude either that there is sufficient evidence that the error rate is low enough that the results deserve to be trusted, or otherwise alert us that errors might be numerous enough to change the results, so the ballots should be examined more carefully. 
Like the election, 
the audit is conducted in such a way that election scrutineers can observe and check that it is being conducted correctly.

The AEC already keeps a careful record that connects each ballot paper to its digital preference list. If there were infinite resources, we could examine this correspondence thoroughly by employing a separate team to compare every single ballot paper with its digital preferences, in the presence of scrutineers. However, this would be much too labor intensive to be practical. Instead, a good approximation can be made by  an audit that takes a random sample of ballot
papers and, for each sampled ballot paper, compares it to its digitised preferences. An \emph{error} is any discrepancy between the digitised preferences and the  marks on the  ballot paper, as interpreted by the people conducting the audit.

Of course, not all errors matter. For example, for the Senate, many preferences are not even counted. Also, errors that advantage the apparent winners usually will not be of concern, nor will errors that affect a candidate who is nowhere close to winning, if they are not too numerous. However, in the Senate count it can be 
very difficult in general to assess whether a group of errors might alter the outcome. Hence we recommend that the audit estimate the average number of errors per ballot
and the percentage of ballots that have at least one error, separately in each state or territory. If the error 
rate is high, suggested ways of dealing with the situation are described below.

An audit should be able to detect errors or malfeasance in any part of the digitisation process. 
It is not appropriate to reuse any part of the software or hardware used for the initial digitisation. A compromise of, or error in, those systems would cause a corresponding error in the audit.

All parts of the digitisation process should be amenable for auditing, in the sense that every possible failure should be detectable, at least with some probability, by the audit process. For example, the audit should be able to detect if a scanner does not produce accurate images, or if the images are altered after they are produced, or if the preference data are not securely stored. This is why it is important to 
commit in advance to the system's output, in the form of digitised preferences, and then to allow scrutineers to observe the retrieval and comparison of the ballot papers. Scrutineers should be encouraged to keep their own records using software of their choice.

Scrutineers need to be able to observe:
    \pbs{add something about seeing custody logs or other evidence that the ballot papers are a complete unadulterated set? VT: They're allowed to watch the whole process - in principle, they're allowed to travel in the cars with the ballots from the polling place. I know you want to write abt ballot custody, logs etc, but I really think it's just a way for us to trip up and get dismissed for now knowing exactly what they already do. That's what the last commissioner was sacked for stuffing up. So I'm writing a general subsection about ballot paper physical security (see below) and *not* writing any detail about what that means.}
\begin{enumerate}
    \item that the audited ballot papers are chosen randomly, after the digitised preferences are committed to;
    \item that the proper ballot papers are retrieved and read;
    \item the written marks on the ballot paper;
    \item how the written marks compare with the digitised preferences.
\end{enumerate}
This sequence of steps shows scrutineers evidence that the audit is being conducted correctly, and allows them to make their own estimate of the error rate if they wish.

\subsection{Ballot paper physical security}
This guide describes an auditing process for ballot papers, but of course this would be meaningless without a secure process for accepting, transporting, and keeping track of ballot papers.  We understand that the AEC already counts first preferences in the polling place, keeps a record of where each ballot box is sent, and (physically) attaches a unique number to each ballot paper, which is also reflected through the electronic process.

These physical security processes are the crucial assumption on which the audit is based and, although we do not detail them here, the audit relies upon them. For example, when a ballot arrives for processing its physical security needs to be established: scrutineers should be able to observe whichever processes the AEC uses, such as verifying physical seals on the ballot box,
reconciling the number of ballots in each batch, checking chain-of-custody documents, logging the examination, etc.

\subsection{Generating trustworthy random ballot selections} \label{subsec:prngs}
The sequence of ballots to be audited needs to be chosen transparently and unpredictably. This is necessary to avoid any (real or perceived) possibility that the sampling might not be truly 
random---unpredictability is necessary to prevent malicious software from hiding misrecordings in ballots that will not be audited. 
Fortunately there are numerous  methods for generating 
pseudorandom sequences of numbers in a transparent, yet unpredictable way. 

The key idea is to use a \emph{random seed} to start a \emph{transparent process}. 
The AEC should commit in advance to the algorithm that will transform the seed into a sequence of ballot selections, and to the method by which that seed will be chosen.
For example, the seed could be derived from a series of dice rolls, or a physical machine such as the one used to draw candidate order on ballot papers. The actual draw should not happen until after the batch has been processed (see below).  It should be generated in public view so that scrutineers can see that it cannot be determined in advance.

Since each state or territory audit is conducted separately, a random seed should be generated separately in each central counting centre, under observation by the scrutineers who are present.

The method for deriving a sequence of ballot selections from the seed should also be published in advance, in the form of openly available computer code for a specific algorithm that 
scrutineers can re-run for themselves. This allows scrutineers to verify that the sequence of ballot selections has been made properly based on the seed.

\clearpage
\section{The steps of an audit}
The Senate election consists of eight simultaneous but separate contests, one for each state and territory. These will each have different candidates and ballots, and potentially different error rates because they are counted using separate equipment. On that basis, we recommend treating each contest  separately.

The precision of the error estimate is related to the sample size in each state or territory. Also,  the sources of error may be specific to a particular counting centre. Therefore, we recommend initially sampling approximately the same number of ballot papers from each state and territory, even though some will have many more ballots than others. This produces an estimate of the error rate in each state or territory, which may be sufficient to conclude the audit in many cases. 

However, in the event of either high error rates or very close outcomes, it may be necessary to take more samples later. This requires completing the digitisation of all the ballots, in order to estimate the closeness of the contest, and then deciding whether a larger sample is required for the necessary level of confidence.

\begin{figure}
\includegraphics[scale=1.75]{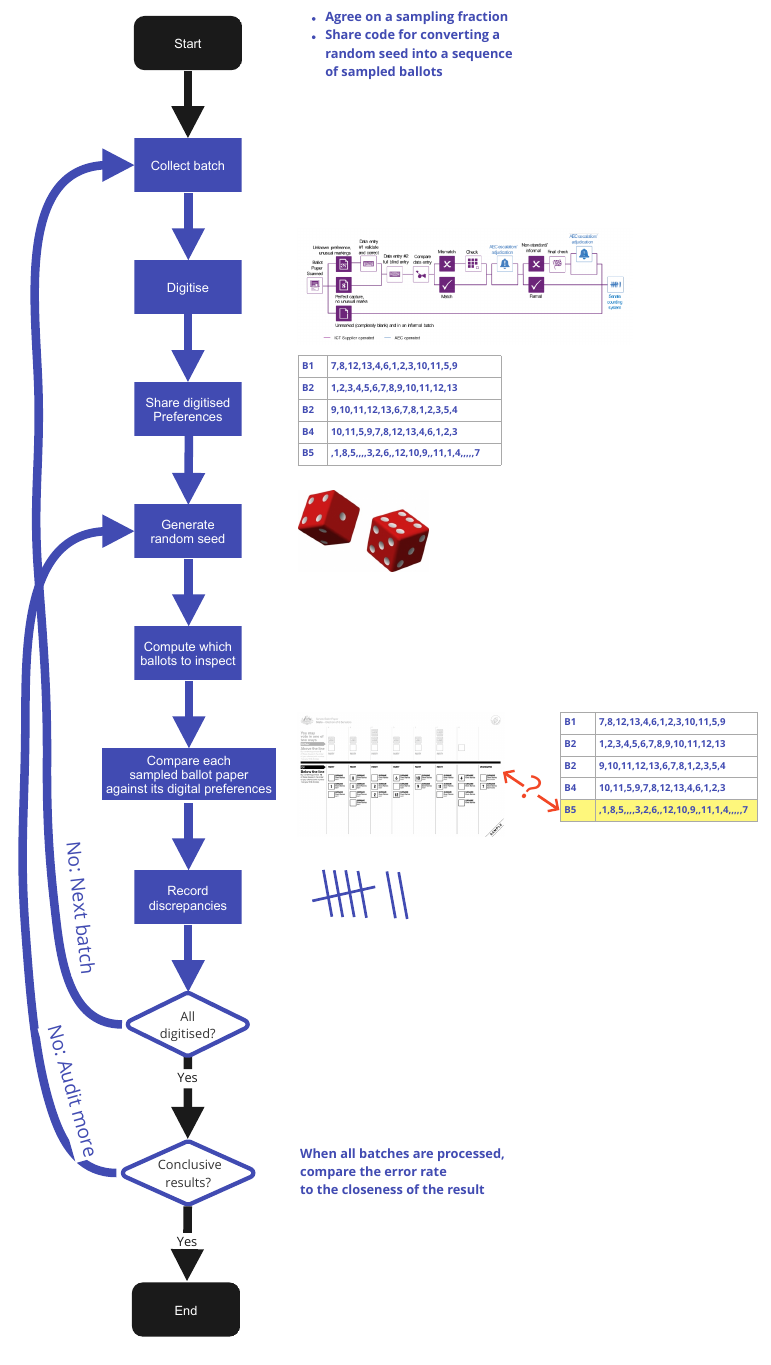}
\caption{Flowchart of Senate auditing process. Image of Senate scanning process from Australian National Audit Office. Image of sample ballot paper from AEC.
\url{https://www.aec.gov.au/voting/how_to_vote/files/senate-btl-sample.png}. Creative Commons License.}
\label{fig:flowchart}
\end{figure}

This section contains an overview of the steps of the audit of the ballot digitisation process, for each state or territory. \autoref{fig:flowchart} shows a diagram of the process. \autoref{sec:details} contains more details about each step.

\subsection{In advance}

In advance, publish the audit methodology, including:
\begin{itemize}
    \item the process for generating a random seed,
    \item the algorithm for deriving a sequence of ballot choices from that seed, including any software that will be used (`sampling software'),
    \item the process for recording errors, including any software that will be used (`auditing software'),
    \item the algorithm for finding confidence bounds on the error rates from the sample,\footnote{The method should be nonparametric, that is, it should not involve assumptions about the distribution of errors.    }
    \item the steps to be taken if the audit shows a high rate of error and more samples need to be taken.
\end{itemize}

\subsection{Preparation on the day}

\begin{itemize}
\item Calculate a sampling fraction based on the threshold in the \textit{Electoral Act 1918}, the expected error rates, desired confidence level, \textit{etc.}, and configure  sampling software accordingly.

\item Report (commit to) the final turnout by polling place before any ballots from a given polling place are processed.
\end{itemize}
Detailed suggestions are in \autoref{subsec:conducting}.

\subsection{Processing and auditing each batch of ballots}

\begin{itemize}
    \item Open the batch, scan it, run it through the digitisation process, generate digitised preferences.
    
    \item Share the digitised preferences with scrutineers, with an index number matching that of the ballot paper.

    \item Generate a random seed in a public ceremony.

    \item Use the sampling software, with the random seed, to determine which ballots to sample from the current batch.

    \item Retrieve these ballots from the batch.

    \item Inspect these ballots and enter the human reading into the auditing software, which records any discrepancies.\pbs{there also need to be official rules for interpreting marks and for resolving ambiguous marks and disagreements among the scrutineers. VT: There are, truly, and they argue about them constantly, and we don't know exactly what they are so I'm sweeping all that under the 'human reading' rug.}
    
    \item Ensure that scrutineers can see what ballots have been selected by the sampling software, can verify that those ballots are the ballots examined, and can compare each to its digitised preferences.
    
\end{itemize}
Detailed suggestions are in \autoref{subsec:conducting}.

\subsection{Analysing the results and concluding the audit}

\begin{itemize}
\item After the final batch, verify that all validly cast ballots have been accounted for and scanned (in a way that scrutineers can check).
\item Use the auditing software to compute and report a confidence interval for the error rate.
\item Make a best-effort computation of the election margin.  
\item If necessary, sample more ballots for any contests in which the audit results are inconclusive.
\item Determine a conclusion and recommendation given the outcomes of the previous steps.
\end{itemize}
Detailed suggestions are in \autoref{subsec:concluding}.

Depending on the exact details of AEC paper-ballot processes, there should also be concluding double-checks to make ensure 
 that every ballot has been through this process and could have been selected to be in the sample by the sampling software.
There should be a reconciliation of numbers, e.g.\ that the number of ballot papers from a given polling place matches the number of eligible voters marked off on the rolls at that place.

\clearpage
\section{Details of each step} \label{sec:details}

\subsection{Before the election: publishing the methodology and process} \label{subsec:InAdvance}

Section 273(AC)(6) states:
\begin{itquote}
Before the polling day for the election, the Electoral Commissioner must publish on the Electoral Commission’s website:
\begin{enumerate}[(a)]
\item the methodology to be used for the ballot paper sampling process; and
\item the process to be used for reconciling preferences.
\end{enumerate}
\end{itquote}
The following sections provide specific suggestions for each of these requirements.

\subsection{Conducting the audit} \label{subsec:conducting}

273(AC)(2) of the \emph{Electoral Act 1918} states:
\begin{itquote}
The Electoral Commissioner must arrange for statistically significant samples of ballot papers to be checked throughout the scrutiny of votes for the election to assure that the electronic data used in counting the votes reflects the data recorded on the ballot papers.
\end{itquote}
Paragraph~(3) requires at least 5,000 ballots to be sampled overall, or 1,000 if only one state's Senate election is running.

It is not specified what ``statistically significant'' means, nor does it explicitly say that the sample must be random, though we take this to be implicit.

This sampling problem is somewhat different from US-style post-election audits because the language ``throughout the scrutiny'' suggests that auditing should be conducted progressively while the ballot digitisation is performed. We assume that ballots are audited in batches, with the audit commencing after the batch has been digitised.
However, it is also important to be able to revisit older batches and examine more ballot papers if concerns arise.

\subsubsection{Generating the sequence of sampled ballots in a transparent way} \label{subsubsec:randGen}
There are many different algorithmic choices for transparent generation of a random sequence of ballots.\footnote{A very simple example based on SHA256 is here: \url{https://www.stat.berkeley.edu/~stark/Java/Html/sha256Rand.htm}} 
We detail one 
particular option here, because it is very simple and all the software for implementing it is already openly available.

Bernoulli sampling using geometric skipping \cite{ottoboni2020},\footnote{Open-source code available at: \url{https://github.com/pbstark/BernoulliBallotPolling/blob/master/code/geometric_skipping.py}}
implements the following idea: from a population of ballots (each with an index number) generate a pseudorandom
sequence of ballot numbers to be audited, in such a way that each ballot is equally likely to be chosen, independently of whether other ballots are chosen. 
Geometric
skipping is conveniently designed to skip forward through a sequence of ballots.\footnote{A useful feature:
if there are big batches, you can restart for every batch and it is still a Bernoulli sample.} 

The geometric skipping algorithm is parameterised by one value: the probability that each  single ballot will be included ($p$). 
The implementation code also accepts a seed value for the pseudorandom number algorithm. 
For Senate auditing, the seed would be generated from some public process such as dice rolling (see \autoref{subsec:prngs}), and the selection probability $p$ would need to be chosen so that the total sample size was compliant with legislation---this is discussed in the 
next section.

\subsubsection{Achieving the required sample size}
\label{sec:required-sample-size}

The \textit{Electoral Act 1918} specifies an overall sample size of at least 5000 ballots, or at least 1000 ballots if only one jurisdiction is running an election at a time.  To start with, we show how to set up the sampling to achieve this minimum requirement.  In \autoref{subsec:concluding} we offer some suggestions for interpreting the `statistically significant' requirement and explain when a larger sample might be required---this would typically happen for very close contests or large error rates. Our suggestion is to make a first pass throughout the digitisation process, sampling ballots at random from each batch as they are digitised. Ideally, this should be sufficient to have confidence in the results in many cases. However, we may discover at the end of the process that the results are inconclusive, and that more auditing is required. In that case, a second pass is needed, retrieving ballots from storage.

We show two sets of worked examples below. The first is where we aim for a sample of at least 1000 for a single jurisdiction. The second is where we aim for a sample of at least 5000 across all jurisdictions. For the latter, we have split the sampling goal equally across the six states and two territories, by setting a target of at least 625 ballots from each one. 

To calculate the required selection probability $p$ for a given a jurisdiction, we need to know the total number of ballots cast in that jurisdiction ($n$) and the target sample size ($t$). The actual sample size ($S$) will follow a binomial distribution ($S \sim \Bi(n, p))$. To ensure that it has at least 99.9\% probability of exceeding the target sample size, we can solve the following inequality in terms of $p$:
\[ \Pr(S \geqslant t) \geqslant 0.999. \]
This can be done using standard statistical software. \autoref{tab:selection-prob} shows some examples, being minimum values of $p$ that satisfy the inequality.

If the actual sample size ends up being smaller than the target sample size (which can occur but is very unlikely), the sample should be supplemented by sampling sufficiently more ballots at random from all of the cast ballots.

There are other reasonable strategies here, for example aiming for a sample of 400 in each jurisdiction and then allowing the remaining 1800 to be sampled from whichever contests are closest or most error prone. These tradeoffs depend on the error rates, how close each contest is, and how much more effort it is to retrieve a ballot that has been put into storage 
than one that has just been digitised. The closeness of the result can be known only after the particular contest is digitised---some past Senate elections have been close enough for fewer than 100 votes to alter the outcome, though tens of thousands is more typical. 

There is limited information about the error rate of the existing process, though of course more information will be available after this year's audit. In prior work, we used a variety of techniques to attempt an estimate, arriving at a value somewhere between 0.04\% and 0.89\% for the per-digit error rate~\cite{blom2020random}. The per-ballot error rate would obviously be some multiple of that, because ballots can have hundreds of digits, though typically have fewer than 10.

\dv{We used \textbf{per-digit} error rates in our earlier paper.  We should carefully distinguish those from per-ballot error rates, which is largely what we discuss in this paper.  I've amended the text above slightly to be accurate about this distinction, but it may require a longer discussion otherwise that detail will get lost on readers.} \todo{Terrific - thanks for fixing. I've added a sentence to emphasise the point.}

\begin{table}
\centering
\caption{\textbf{Example minimum values for the selection probability, $p$.}  The calculations assume all enrolled voters cast a vote. Enrolment numbers are from the AEC, \url{https://www.aec.gov.au/Enrolling_to_vote/Enrolment_stats/national/2021.htm}.}
\smallskip
\begin{tabular}{lrcc}
\toprule
 &  & \multicolumn{2}{c}{Selection probability ($p$)} \\
\cmidrule{3-4}
State/Territory & Enrolled voters ($n$) &
Target: $t = 1000$ & Target: $t = 625$ \\
\midrule
NSW &	5,427,292 & 0.000202852 & 0.000129934 \\
VIC &	4,305,961 &	0.000255677 & 0.000163770 \\ 
QLD &	3,450,635 & 0.000319053 & 0.000204365 \\
WA 	&   1,752,273 & 0.000628288 & 0.000402441 \\
SA 	&   1,244,611 & 0.000884557 & 0.000566591 \\ 
TAS &	  397,279 & 0.002771133 & 0.001775011 \\
ACT & 	  309,521 &	0.003556806 & 0.002278264 \\
NT 	&     145,335 &	0.007574710 & 0.004851881 \\
\bottomrule
\end{tabular}
\label{tab:selection-prob}
\end{table}

So another reasonable approach is to make a guess at the error rate and the likely closeness of the result, and to compute a sample target that would give us sufficient confidence
in the result, if those guesses turned out to be correct. 
For example, if we assume an error rate of 0.5\% and that approximately 1\% of votes would need to change in order to change the election result, we could use the techniques described in
\autoref{subsubsec:confidenceintervals} to choose a sample size so that, if those assumptions were correct, the audit could conclude with confidence that the error rate was low enough, without needing a second pass.
 This may require a larger first-pass sample, but might be worth the effort to avoid additional passes, which involve time-consuming retrieval from storage. 
 Of course, if the assumptions are shown to be wrong, a second pass may be necessary anyway.

In summary, depending on the relative difficulty of auditing a ballot that has just been digitised vs auditing that same ballot after it has been stored, there are several options:
\begin{itemize}
    \item target exactly the sample size needed to meet the legislative requirement,
    \item target a slightly larger sample size, hoping to reduce the number of states or territories that require additional passes,
    \item target a smaller sample size, expecting to use all of the extra on the one or two states or territories that require additional passes,
    \item use an estimate of the error rate to choose a sample size that is likely to be large enough to avoid the need for additional passes.
\end{itemize}

The following discussion is independent of which of these options is chosen.

\subsection{Dealing with discrepancies}
\label{subsec:concluding}

Software-based processes are human processes, and humans are imperfect and make mistakes. It is possible that the audit will uncover some discrepancies that are enough to alter the outcome, for example because of a systematic error in the processing or a human error in the data uploading. In these cases, it is important to be honest and transparent, and to examine what went wrong while allowing scrutineers to observe the reconciliation process. The legislation does not specify what to do in the event of larger-than-expected error rates, but one obvious option is to audit more so as to get a 
better estimate. If the error rate seems to be genuinely high, manual examination of all the paper ballots may be required.

The legislation also does not specify what level of error is small enough to be acceptable. In general, tighter contests will be less tolerant of errors, while larger errors may make no difference if the margin is sufficiently large.

The rest of this section explains how to compare the estimated error rate with calculations of how many vote-changes can alter the election outcome.

\subsubsection{Estimating the overall error rate} \label{subsubsec:confidenceintervals}

Once the sample of ballots is collected, they are compared to the digitised preferences and the number of errors are recorded.  This allows us to estimate the overall error rate.  For example, if we observe 30 errors out of 5000 ballots, an estimate of the error rate is $30 / 5000 = 0.006 = 0.6\%$.

In general, this will be close but not equal to the true error rate.  To represent the uncertainty of the estimate we can calculate a confidence interval.  This will give a range of values (for the true error rate) that are most plausible given the data.  For the above example, a 95\% confidence interval is (0.0041, 0.0086).\footnote{Using the Clopper--Pearson method, which assumes the sample size of 5,000 is fixed in advance.}
\pbs{This isn't quite right. Clopper-Pearson is for binomial, not bernoulli.}
\dv{It's for a fixed number of Bernoulli trials, but I agree that would be confusing here because it's not the same as Bernoulli sampling.  I think we can just avoid referring to either one, I'll shorten the description above.}
In other words, it is implausible that the true error rate is outside the range 0.41\% to 0.86\%.\footnote{`Implausible' is a matter of degree.  It is possible that the true value is outside the range---we cannot rule that out with complete certainty with only a sample---but it is more plausible that it is within the range.}

The legislation refers to a ``statistically significant sample,'' 
without defining what this means.  
In the current context, a typical interpretation of ``statistically significant'' would be that the confidence interval excludes some comparison value of interest; we adopt this interpretation in the following sections, where we compare the interval to an estimate of how close the election was (a `margin').  
Therefore, we interpret ``statistically significant sample'' to refer to a sample that allows us to estimate the error precisely enough---by producing a narrow enough confidence interval---that it excludes the (margin) value we are comparing it to.  
The main factor that will determine this is the number of sampled ballots: sampling more ballots leads to narrower confidence intervals.  
If the initial sample does not give a statistically significant result, we advocate sampling more ballots until statistical significance is achieved (see \autoref{sec:concluding-the-audit}), as required by the legislation or, if that is not possible, manually re-examining all the ballots to ensure an accurate result.

The `95\%' mentioned above is known as the \emph{confidence level} of the confidence interval.  
It tell us how reliable the interval is---the higher the confidence level, the more often the confidence interval will include the true value.\footnote{More precisely, it is a probability under the sampling procedure.  
If we were to obtain all possible such samples, and for each one calculate a confidence interval, at least 95\% of the intervals would contain the true value.}
We can set the confidence level to any value between zero and 100\%, but there is a tradeoff: all else equal, to increase the level requires making the confidence interval wider, which would then require more samples to reach statistical significance.\footnote{Thus, the concept of `statistical significance' is always with reference to a particular choice of confidence level.  For transparency, the confidence level used should always be reported.}  A conventional choice through much of scientific research is 95\%, and in some contexts it is much higher, but rarely lower.

There are many methods for calculating confidence intervals.  They differ in the assumptions they make about how the data are sampled.  One should use a method that is consistent with the sampling design
and what is known about the population being sampled.  
The above example used the Clopper--Pearson method, which is widely used and implemented in standard software.  
The method assumes that the population is binary (each ballot either does or does not have an error), and that the sample size is fixed in advance and drawn  `with replacement', which is appropriate for many studies.  
In the current context, ballots will actually be sampled `without replacement' and the size is not fixed in advance; methods exist that are appropriate for such samples, see \cite{ottoboni2020}.
Another difference is that we may need to do more than one round of sampling to obtain a statistically significant estimate.  
If so, the calculation of the confidence intervals needs to be adapted to take this into account.  This requires more advanced methods.\footnote{%
If there will be only one round of sampling, the Clopper--Pearson confidence interval for the Binomial is valid, conditional on the attained sample size.
If one wants to allow additional rounds of sampling at will, there are several options:
a sequential method can be used
after randomly permuting the data, for instance, \cite{howardEtal21} or
\cite{waudby-smithEtal21}.
If one wants to limit the number of rounds of sampling, e.g., to at most two before a full manual inspection will be performed, Clopper-Pearson could be used at each stage, with the confidence level for each round adjusted (e.g., using Bonferroni's inequality) so that the overall confidence level of the procedure remains at least 95\%.
}

\subsubsection{What size error is small enough?}
\label{subsubsec:whatErrorIsSmallEnough}

In first-past-the-post elections, where the winner is simply the person who receives the most votes, the error rate can simply be compared to the \emph{margin of victory} (or just \emph{margin}), that is the number of ballots that would need to change to change the result. 
In first-past-the-post elections the margin of victory is half the difference between the tally of the winner and the second highest tally. 
For example, if 45\% of voters reportedly voted for Alice and 35\% reportedly voted for Bob, the audit needs to provide confidence that errors affected less than 5\% of the ballots. 
For these simple electoral systems, even more careful accounting is possible---errors that advantage Bob can be disregarded, for example, because they would not change the election outcome. There is an extensive literature on precise, efficient, accounting for different kinds of errors~\cite{lindeman2012gentle,stark2020sets}.

Unfortunately, none of these techniques transfers easily to Senate auditing, because it is infeasible to compute election margins, and just as difficult to tell which kinds of errors could make a difference to the outcome. Even if a certain preference was not read during the official count, an alteration in some other ballot may change the elimination order and cause that preference to matter.

However, the error estimation is still valuable. It can provide a good degree of confidence under reasonable assumptions in many scenarios, or, conversely, it can give a meaningful indication that more examination is needed, if the error rate is high.

Although computing exact margins for the Senate count is infeasible in general, there are some straightforward ways to find small changes that are sufficient to change the outcome. 
For example, in many cases the last seats are allocated when a candidate ($c$) is eliminated, leaving exactly as many continuing candidates as unfilled seats. In this case, it is easy to compute the difference between $c$ and the next-lowest candidate $d$ (who wins a seat). Shifting sufficient votes from $d$ to $c$ obviously changes the outcome---$c$ wins a seat instead of $d$. The required number of votes is called the \emph{last-round margin}. At any time, the outcome can also be changed by raising up any non-winning candidate until they have a quota.

However, there may be smaller alterations earlier in the elimination sequence that are also sufficient to change the outcome. These are much harder to detect in general, but there are several published heuristic algorithms for finding answers~\cite{blom2020did, iVoteLGE}. These programs search for small changes in votes that  change the election outcome. 
If a program finds an example, that change to the votes definitely can change the outcome---but there may be a way to change the outcome by changing fewer votes that the algorithm did not detect.
Open-source software for these algorithms are available and could be run routinely on all digitised Senate preferences.

Any of these methods gives a feasible, but not necessarily minimal, way of changing the outcome. 
We call this number of vote-changes the \emph{apparent margin}. 
Evidence that there were fewer errors than the apparent margin is evidence against that particular change in the outcome.

\subsubsection{Update (22 June 2022): results for the 2022 Australian Senate election} \label{subsubsec:2022Data}
We have applied some heuristic search algorithms to the openly available vote data supplied by the AEC.
Because first preferences are manually counted in the polling place in the presence of scrutineers, the opportunity for them to be altered undetectably by software problems is fairly limited. We therefore give two versions of the smallest-change values: one that allows for first preferences to be altered, and one that is limited to changes in later preferences only.
Full results are in Table~\ref{tab:smallChanges}. Instructions for replicating the results are in the source code repository.\footnote{\url{https://github.com/AndrewConway/ConcreteSTV/blob/main/reports/Margins2022.pdf}}

\begin{table}[h]
\noindent
\begin{tabular}{lrrrrrl}
\textbf{State or} & \textbf{Formal}  & \multicolumn{2}{c}{\textbf{Allow 1st-prefs}} & \multicolumn{2}{c}{\textbf{No 1st-prefs}} &  \textbf{Effect} \\
\textbf{Territory} & \textbf{votes}& \textbf{Change} &  \textbf{(\%)} & \textbf{Change} & \textbf{(\%)} &  \\

ACT&285217&14137&4.96\%&$\times$&$\times$&+Seselja(LP); -Pocock(Ind)\\
NSW&4800722&57340&1.19\%&57340&1.19\%&+McCulloch(ON); -Molan(LP)\\
NT&103617&11412&11.01\%&$\times$&$\times$&+Anlezark(Grn); -McCarthy(ALP)\\
QLD&3013868&54810&1.82\%&59428&1.97\%&+Stoker(LP); \\
&&&&&&-Chisolm (ALP) (1st pref changes) \\
&&&&&& -Hanson(ON) (no 1st prefs)\\
SA&1128524&9306&0.82\%&12558&1.11\%&+Gill(ALP); -Liddle(LP)\\
TAS&361048&11697&3.24\%&$\times$&$\times$&+Mav(ON); -Tyrrell(JL)\\
VIC&3821539&9341&0.24\%&9341&0.24\%&+Pickering(ON); -Babet(UAP)\\
WA&1526123&11745&0.77\%&$\times$&$\times$&+Filing (ON); -Payman (ALP)\\
\end{tabular} \caption{Vote changes that can change Senate outcomes for 2022 Australian Senate results. Two kinds of changes are shown. `Allow 1st-prefs' lets first-preferences be different. In `No 1st-prefs', first-preferences are assumed to be perfectly recorded. In some cases, this results in a slightly larger number of vote changes, in others it makes no difference, and in some cases it precludes any solution---this is indicated with a '$\times$'.}\label{tab:smallChanges}
\end{table}

Victoria seems to be the closest, with 9341 vote changes, or $0.24\%$, sufficient to change the outcome. Western Australia (0.77\%) and South Australia (0.82\%) are next. The other states do not seem close---the smallest changes we could find all required more than a $1\%$ vote change. It would therefore make sense to concentrate auditing effort on Victoria, WA and SA. If the error rate is indeed less than $0.5\%$, then we can be fairly confident that the WA and SA results are correct, but it may take a fairly large sample to have confidence that the error rate is sufficiently small. In the case of Victoria, based on previous AEC statements, it is possible that the overall error rate may be larger than the number of  changes sufficient to change the outcome. This does not imply that the results are wrong, but it does mean that more extensive re-examination of the ballot papers may be required in order to be confident of the result.

\subsection{Concluding the audit}
\label{sec:concluding-the-audit}

After all the votes have been processed and each batch has been audited, we have both a complete list of digitised preferences and an overall estimate (and confidence interval) of the error rate in each state or territory.  In addition, for each state and territory we have an apparent margin: an estimate of the number of ballots that would have to have digitisation errors for the reported electoral outcome to be incorrect.

\begin{figure}
\centering
\includegraphics[width=0.8\textwidth]{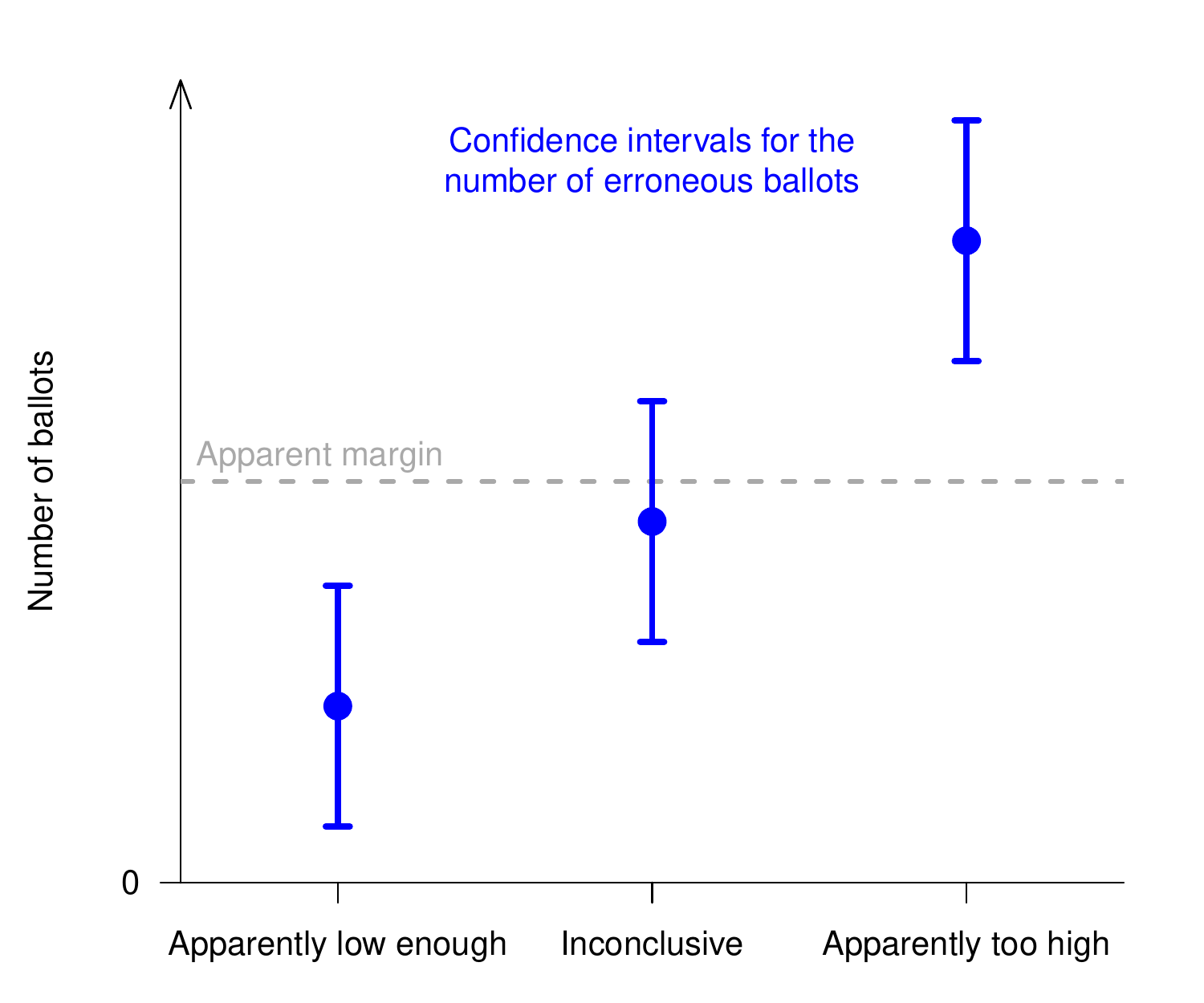}
\caption{Possible outcome scenarios.}
\label{fig:conclusions-plot}
\end{figure}

We can multiply the confidence intervals by the number of cast ballots, so they become confidence intervals for the \emph{number} of ballots with digitisation errors.  This allows us to compare them directly with the apparent margins.
In each jurisdiction we have one of the following possible scenarios, as illustrated in \autoref{fig:conclusions-plot}:
\begin{itemize}
    \item \textbf{Apparently low enough error rate.} The upper endpoint of the confidence interval is below the apparent margin. In this case, it is appropriate to conclude the audit and report that the errors are unlikely to have altered the outcome of the election. 
    (However, the apparent margin may be an overestimate; this should be noted as a caveat in the report.)

    \item \textbf{Apparently too high error rate.} The lower endpoint of the confidence interval is larger than the apparent margin. This is the difficult case, which may indicate irrelevant errors or a genuine problem: it requires further investigation.  
    \begin{itemize}
        \item Assess the nature of the errors.
        \item Investigate potential sources of error.  For example, this might lead to an observation that one scanner is of very poor quality, or one computer involved in the digitisation process has been compromised.
    \end{itemize}
    There is no general procedure for dealing with this situation. It requires careful assessment of the nature and causes of the errors and, potentially, re-examination of all the paper ballots.

    \item \textbf{Inconclusive error rate.} The confidence interval includes the apparent margin. In this case, we have insufficient evidence to say whether the error rate is greater or smaller than the margin; in statistical jargon, the difference between them is said to be `not statistically significant'.  
    
    In this case, more auditing should be done to get a more precise estimate of the error rate, to determine with more certainty whether or not it exceeds the margin. This second phase should ensure that all ballots cast in the contest could be sampled with equal probability.
\end{itemize}
In addition to meeting the legislative requirements, these investigations and outcomes could also be used to defend an election outcome against an accusation that
the result is not accurate. A carefully conducted, rigorous demonstration of a low error rate should be convincing evidence that the outcome is correct.
A candidate who disputed the results could be invited to produce a specific set of ballot alterations that could allow them to be elected, and if they were unable to produce any within the measured error rate this would be strong evidence against that particular set of errors.

\section{Conclusion}
Overall, a transparent audit should improve both the real and perceived accuracy and integrity of Australian Senate elections. 
If done properly, it has the potential either to detect and correct errors, or to demonstrate transparently that the error rate is low enough to 
support an accurate election result.

\newpage
\bibliographystyle{alpha}
\bibliography{sample}

\end{document}